\documentclass[10pt]{article}

\usepackage{times}
\usepackage{amssymb,amsmath,amsthm}
\usepackage{mathrsfs}

\DeclareMathAlphabet\EuScript{U}{eus}{m}{n}
\SetMathAlphabet\EuScript{bold}{U}{eus}{b}{n}

\newcommand{\eg}{\emph{e.g.}~}

\def\B{\mathscr{B}}
\def\C{\mathbb{C}}
\def\D{\mathscr D}
\def\E{\mathscr E}
\def\F{\mathscr F}
\def\G{\mathscr G}
\def\H{\mathcal H}
\def\K{\mathcal K}
\def\N{\mathbb{N}}
\def\P{\mathcal P}
\def\R{\mathbb{R}}
\def\T{\mathcal T}
\def\U{\mathscr U}

\def\<{\left\langle}
\def\>{\right\rangle}
\def\({\left(}
\def\){\right)}
\def\[{\left[}
\def\]{\right]}
\def\lone{\mathsf{L}^{\:\!\!1}}
\def\loneloc{\mathsf{L}^{\:\!\!1}_{\rm loc}}
\def\ltwo{\mathsf{L}^{\:\!\!2}}
\def\linf{\mathsf{L}^{\:\!\!\infty}}
\def\e{\mathrm{e}}
\def\de{\mathrm{d}}
\def\dom{\matheucal D}

\newtheorem{Theorem}{Theorem}[section]
\newtheorem{Proposition}[Theorem]{Proposition}
\newtheorem{Lemma}[Theorem]{Lemma}
\newtheorem{Remark}[Theorem]{Remark}
\newtheorem{Corollary}[Theorem]{Corollary}
\newtheorem{Assumption}[Theorem]{Assumption}
\newtheorem{Definition}[Theorem]{Definition}

\begin{document}
  
  
  \title{{\Large\textbf{Time delay and short-range scattering in quantum waveguides}}}
  
  \author{Rafael Tiedra de Aldecoa}
  \date{\small
    \begin{quote}
      \emph{
    \begin{itemize}
    \item[]
      D\'epartement de physique th\'eorique, Universit\'e de Gen\`eve,\\
      24, quai E. Ansermet, 1211 Gen\`eve 4, Switzerland
    \item[]
      \emph{E-mail\:\!:}
      rafael.tiedra@physics.unige.ch
    \end{itemize}
      }
    \end{quote}
  }
  \maketitle

  
  \begin{abstract}
    Although many physical arguments account for using a modified definition of time delay in
    multichannel-type scattering processes, one can hardly find rigorous results on that issue in
    the literature. We try to fill in this gap by showing, both in an abstract setting and in a
    short-range case, the identity of the modified time delay and the Eisenbud-Wigner time delay in
    waveguides. In the short-range case we also obtain limiting absorption principles, state spectral
    properties of the total Hamiltonian, prove the existence of the wave operators and show an
    explicit formula for the $S$-matrix. The proofs rely on stationary and commutator methods.
  \end{abstract}
  
  \section{Introduction and main results}
  \setcounter{equation}{0}
  
  This paper is concerned with time delay (defined in terms of sojourn times) in scattering theory
  for waveguides. Our main aim is to show that, as in $N$-body scattering and scattering by step
  potentials, one has to use a modified definition of time delay in order to prove its existence and
  its identity with the Eisenbud-Wigner time delay. We refer to \cite{Martin75} for the treatment of
  this issue in the case of scattering with dissipative interactions.

  Let us first recall the standard definition \cite{JSM} of time delay for an elastic two-body
  scattering process. Given a free Hamiltonian $H_0$ and a total Hamiltonian $H$ such that the wave
  operators $W^\pm$ exist and are complete, one defines for certain states $\varphi$ and $r>0$ two
  sojourn times, namely:
  \begin{equation}\label{free elastic sojourn}
    T^0_r(\varphi):=\int_{-\infty}^\infty\de t\int_{|x|\leq r}\de^3x
    \left|(\e^{-itH_0}\varphi)(x)\right|^2
  \end{equation}
  and
  \begin{equation}\label{full elastic sojourn}
    T_r(\varphi):=\int_{-\infty}^\infty\de t\int_{|x|\leq r}\de^3x
    \left|(\e^{-itH}W^-\varphi)(x)\right|^2.
  \end{equation}
  The first number is interpreted as the time spent by the freely evolving state $\e^{-itH_0}\varphi$
  inside the ball $\mathcal B_r:=\{x\in\R^3\:\!:\:\!|x|\leq r\}$, whereas the second one is
  interpreted as the time spent by the associated scattering state $\e^{-itH}W^-\varphi$ within the
  same region. Since $\e^{-itH}W^-\varphi$ is asymptotically equal to $\e^{-itH_0}\varphi$ as
  $t\to-\infty$, the difference
  \begin{equation*}
    \tau_r^{\rm in}(\varphi):=T_r(\varphi)-T^0_r(\varphi)
  \end{equation*}
  corresponds to the time delay of the scattering process with incoming state $\varphi$ for the ball
  $\mathcal B_r$. The (global) time delay of the scattering process with incoming state $\varphi$ is,
  if it exists, the limit of $\tau_r^{\rm in}(\varphi)$ as $r\to\infty$. For a suitable initial state
  $\varphi$ and a sufficiently short-ranged interaction, it is known \cite{Amrein/Cibils,ACS} that
  this limit exists and is equal to the expectation value in the state $\varphi$ of the
  Eisenbud-Wigner time delay operator.

  If the scattering process associated to the pair $\{H_0,H\}$ is inelastic (typically of a $N$-body
  nature), then one has to modify the definition of time delay. The heuristic argument goes as follows.
  Due to the inelastic nature of the interaction, the expectation values of the momentum operator in
  the state $\e^{-itH}W^-\varphi$ and in the state $\e^{-itH_0}\varphi$ may converge to different
  constants as $t\to+\infty$. This would result in the divergence of the retardation (or advance) of
  the state $\e^{-itH}W^-\varphi$ with respect to the state $\e^{-itH_0}\varphi$. Similarly, if the
  incoming state $\varphi$ is replaced by the outcoming state $S\varphi$, where $S$ is the scattering
  operator, then the same divergence, but with an opposite sign, would occur as $t\to-\infty$.
  Therefore, in order to cancel both divergences out, $T_r(\varphi)$ should not be compared with the
  free sojourn time $T^0_r(\varphi)$, but with an effective free sojourn time involving both
  $T^0_r(\varphi)$ and $T^0_r(S\varphi)$. A symmetry argument \cite[Sec.~V.(a)]{Martin81} leads
  naturally to the mean value $\frac12\[T^0_r(\varphi)+T^0_r(S\varphi)\]$ for this effective time.
  Thus one ends up with the expression
  \begin{equation}\label{delay}
    \tau_r(\varphi):=T_r(\varphi)-\mbox{$\frac12$}\[T^0_r(\varphi)+T^0_r(S\varphi)\]
  \end{equation}
  for the time delay of the inelastic scattering process with incoming state $\varphi$ for the ball
  $\mathcal B_r$. In the case of $N$-body scattering and step potential scattering, one can easily
  generalize the definition (\ref{delay}) to its multichannel counterpart
  \cite{Smith60,Bolle/Osborn,Martin81}.

  Now consider a waveguide $\Omega:=\Sigma\times\R$ with coordinates $(x',x)$, where $\Sigma$ is a
  bounded open connected set in $\R^{d-1}$, $d\geq2$. Let $H_0:=-\Delta_{\rm D}^\Omega$ be the
  Dirichlet Laplacian in $\ltwo(\Omega)$ (equipped with the norm $\|\cdot\|$). Let $H$ be a
  selfadjoint perturbation of $H_0$ such that the wave operators
  $W^\pm:=\textrm{s-}\lim_{t\to\pm\infty}\e^{itH}\e^{-itH_0}$ exist and are complete (so that the
  scattering operator $S:=(W^+)^*W^-$ is unitary). Then the associated scattering process is globally
  elastic, but the kinetic energy along the $x$-axis is not conserved if the interaction is general
  enough. On the other hand, the waveguide counterparts of the sojourn times
  (\ref{free elastic sojourn}) and (\ref{full elastic sojourn}) must be
  \begin{equation}\label{free sojourn}
    T^0_r(\varphi):=\int_{-\infty}^\infty\de t\left\|F_r\e^{-itH_0}\varphi\right\|^2
  \end{equation}
  and
  \begin{equation}\label{full sojourn}
    T_r(\varphi):=\int_{-\infty}^\infty\de t\left\|F_r\e^{-itH}W^-\varphi\right\|^2,
  \end{equation}
  where $F_r$ denotes the projection onto the set of the states localized in the cylinder
  $\Omega_r:=\Sigma\times[-r,r]$. Thus the sojourn times involve regions expanding in the
  $x$-direction, the axis along which the scattering process is inelastic. This explains why we have
  to use the formula (\ref{delay}) when defining time delay in waveguides. As in the $N$-body case,
  one can also write the time delay given by (\ref{delay})--(\ref{full sojourn}) in a multichannel
  way (see Remark \ref{multichannel}).

  Let us fix the notations and recall some properties of $H_0$ before giving a description of our
  results. $\otimes$ (resp. $\odot$) stands for the closed (resp. algebraic) tensor product of
  Hilbert spaces or of operators. Given two Hilbert spaces $\H_1$ and $\H_2$, we write
  $\H_1\subset\H_2$ if $\H_1$ is continuously embedded in $\H_2$ and $\H_1\simeq\H_2$ if $\H_1$ and
  $\H_2$ are isometric. $\B(\H_1,\H_2)$ stands for the set of bounded operators from $\H_1$ to $\H_2$
  with norm $\|\cdot\|_{\H_1\to\H_2}$, and $\B(\H_1):=\B(\H_1,\H_1)$. $\|\cdot\|$ (resp.
  $\<\:\!\cdot\:\!,\:\!\cdot\:\!\>$) denotes the norm (resp. scalar product) of the Hilbert space
  $\H:=\ltwo(\Omega)\simeq\ltwo(\Sigma)\otimes\ltwo(\R)$. If there is no risk of confusion, the
  notations $\|\cdot\|$ and $\<\:\!\cdot\:\!,\:\!\cdot\:\!\>$ are also used for other spaces. $Q$
  (resp. $P$) stands for the position (resp. momentum) operator in $\ltwo(\R)$. $\N:=\{0,1,2,\ldots\}$
  is the set of natural numbers. $\H^k(\Sigma)$, $k\in\N$, are the usual Sobolev spaces over $\Sigma$,
  and $\H^s_t(\R^n)$, $s,t\in\R$, $n\in\N\setminus\{0\}$, are the weighted Sobolev spaces over $\R^n$
  \cite[Sec.~4.1]{ABG} (with the convention that $\H^s(\R^n):=\H^s_0(\R^n)$ and
  $\H_t(\R^n):=\H^0_t(\R^n)$). Given a selfadjoint operator $A$ in a Hilbert space $\H$, we write
  $E^A(\:\!\cdot\:\!)$ for the spectral measure of $A$ and $\dom(A)$ for the domain of $A$ endowed
  with its natural graph topology. $\chi_{[-r,r]}$ is the characteristic function for the interval
  $[-r,r]$ and $\<\:\!\cdot\:\!\>:=\sqrt{1+|\cdot|^2}$.

  The Dirichlet Laplacian $-\Delta_{\rm D}^\Sigma$ in $\ltwo(\Sigma)$ has a purely discrete spectrum
  $\T:=\{\nu_\alpha\}_{\alpha\geq1}$ consisting of eigenvalues $0<\nu_1<\nu_2\leq\nu_3\leq\ldots$
  repeated according to multiplicity. In particular $-\Delta_{\rm D}^\Sigma$ admits the spectral
  decomposition $-\Delta_{\rm D}^\Sigma=\sum_{\alpha\geq1}\nu_\alpha\P_\alpha$, where $\P_\alpha$ is
  the one-dimensional orthogonal projection associated to $\nu_\alpha$. The Dirichlet Laplacian
  $-\Delta_{\rm D}^\Omega$ can be written as
  $-\Delta_{\rm D}^\Omega=-\Delta_{\rm D}^\Sigma\otimes1+1\otimes P^2$, so that $H_0$ has a purely
  absolutely continuous spectrum coinciding with the interval $[\nu_1,\infty)$. Since $S$ commutes
  with $H_0$, $S$ can be expressed as a direct integral of unitary operators $S(\lambda)$,
  $\lambda\geq\nu_1$, where $S(\lambda)$ acts in the fiber at energy $\lambda$ for the spectral
  decomposition of $H_0$ (see Section \ref{spectral decomposition}). $S(\lambda)$ is called the
  $S$-matrix at energy $\lambda$.

  \begin{Definition}
    Let $\sigma_{\rm p}(H)$ be the set of eigenvalues of $H$ and $t\geq0$, then
    \begin{align*}
      \D^\Omega_t&:=\left\{\varphi\in\ltwo(\Sigma)\otimes\H_t(\R)\:\!:\:\!
      E^{H_0}(J)\varphi=\varphi~\textrm{for some compact set}~J~\textrm{in}~
      (\nu_1,\infty)\setminus\(\sigma_{\rm p}(H)\cup\T\)\right\},\\
	\D^\R_t&:=\big\{\varphi\in\H_t(\R)\:\!:\:\!E^{P^2}(J)\varphi=\varphi~
	\textrm{for some compact set}~J~\textrm{in}~\R\setminus\{0\}\big\}\:\!.
    \end{align*}
  \end{Definition}

  It is clear that $\D^\R_t$ is dense in $\ltwo(\R)$ and that $\D^\R_{t_1}\subset\D^\R_{t_2}$ if
  $t_1\geq t_2$. The spaces $\D^\Omega_t$ also satisfy $\D^\Omega_{t_1}\subset\D^\Omega_{t_2}$ if
  $t_1\geq t_2$, and $\D^\Omega_t$ is dense in $\H$.

  We are in a position to state our results. In Section \ref{abstract existence}, we prove the
  following general existence criterion. It involves the Eisenbud-Wigner time delay operator
  $\tau_\textsc{e-w}$, which is the decomposable operator in the spectral decomposition
  of $H_0$ formally defined by the family
  \begin{equation*}
    \tau_\textsc{e-w}(\lambda)
    :=-iS(\lambda)^*\:\!\frac{\de S(\lambda)}{\de\lambda}\:\!,\quad\lambda\geq\nu_1\:\!.
  \end{equation*}
    
  \begin{Theorem}\label{abstract time delay}
    Let \:\!$\Omega:=\Sigma\times\R$, where $\Sigma$ is a bounded open connected set in
    $\R^{d-1}$, $d\geq2$. Consider a (two-body) scattering system in the Hilbert space
    $\H:=\ltwo(\Omega)$ with free Hamiltonian $H_0:=-\Delta^{\Omega}_{\rm D}$ and total Hamiltonian
    $H$. Suppose that
    \begin{enumerate}
    \item For each $r>0$ the projection $F_r$ is locally $H$-smooth on
      $\,(\nu_1,\infty)\setminus\(\sigma_{\rm p}(H)\cup\T\)$.
    \item The wave operators $W^\pm$ exist and are complete.
    \end{enumerate}
    Let $\varphi\in\D^\Omega_2$ be such that $S\varphi\in\D^\Omega_2$ and
    \begin{equation*}
      \left\|\(W^--1\)\e^{-itH_0}\varphi\right\|\in\lone((-\infty,0),\de t)
    \end{equation*}
    and
    \begin{equation*}
      \left\|\(W^+-1\)\e^{-itH_0}S\varphi\right\|\in\lone((0,\infty),\de t)\:\!.
    \end{equation*}
    Then $\tau_r(\varphi)$ exists for each $r>0$ and $\tau_r(\varphi)$ converges as $r\to\infty$ to
    a finite limit. If in addition the function $\lambda\mapsto S(\lambda)$ is strongly continuously
    differentiable on an open set $J\subset(\nu_1,\infty)$ such that $E^{H_0}(J)\varphi=\varphi$,
    then $\lim_{r\to\infty}\tau_r(\varphi)=\<\varphi,\tau_\textsc{e-w}\varphi\>$.
  \end{Theorem}

  Using the stationary formalism of \cite{Kuroda73} and the commutator methods of \cite{ABG}, we show
  in Section \ref{short-range scattering} some results concerning short-range scattering theory in
  waveguides. In Theorem \ref{limiting absorption}, we obtain limiting absorption principles (which
  lead to the existence of the wave operators) and state spectral properties of the total
  Hamiltonian. We also prove a result on the norm differentiability of the $S$-matrix (Proposition
  \ref{S diff}) which relies on an explicit formula for the $S$-matrix (Lemma \ref{S formula}). In
  Section \ref{short-range existence}, we use the results of Section \ref{short-range scattering} to
  find sufficient conditions under which the hypotheses of Theorem \ref{abstract time delay} are
  satisfied (see Theorem \ref{short-range theorem} for the precise statement):

  \begin{Theorem}
    Let $H:=H_0+V$, where $V$ decays as $|x|^{-\kappa}$, $\kappa>4$, at infinity. Then there exists
    a dense set $\E$ such that, for each $\varphi\in\E$, $\tau_r(\varphi)$ exists for all $r>0$ and
    $\tau_r(\varphi)$ converges as $r\to\infty$ to a finite limit equal to
    $\<\varphi,\tau_\textsc{e-w}\varphi\>$.
  \end{Theorem}

  \begin{Remark}
    A comparison with the corresponding theorem \cite[Prop.~4]{ACS} for scattering in $\R^d$, shows
    us that potentials decaying as $|x|^{-\kappa}$, $\kappa>2$, at infinity may also be treated.
    This could certainly be done by adapting results on the mapping properties of the scattering
    operator (\eg \cite{ACS,Jensen/Nakamura}) to the waveguide case. However, since these properties
    deserve a study on their own, we prefer not to use them in the present paper.
  \end{Remark}
  
  We finally mention Lemma \ref{diff. of trace} which establishes some regularity properties of the
  trace-type operator associated to the spectral transformation for $H_0$.

  \section{General existence of time delay in waveguides}
  \setcounter{equation}{0}

  \subsection{Preliminaries}

  In the sequel we give sufficient conditions for the existence of the time delay in $\Omega_r$. Then
  we show that the (global) time delay, if it exists, is expressed in terms of the limit of an
  auxiliary time. We start by recalling some facts which will be freely used throughout the paper.

  The one-dimensional Fourier transform $\F$ is a topological isomorphism of $\H^s_t(\R)$ onto
  $\H^t_s(\R)$ for any $s,t\in\R$. Given two separable Hilbert spaces $\H_1$ and
  $\H_2$ one has the relation $(\H_1\otimes\H_2)^*\simeq\H_1^*\otimes\H_2^*$ for their adjoint
  spaces. Furthermore, if $1$ is the identity operator in $\H_1$ and $A$ a selfadjoint operator in
  $\H_2$, then one has the identity $\dom(1\otimes A)\simeq\H_1\otimes\dom(A)$. If $\H_1$, $\H_2$,
  $\K_1$, $\K_2$ are Hilbert spaces and $A_i\in\B(\H_i,\K_i)$ ($i=1,2$), then
  $A_1\otimes A_2\in\B(\H_1\otimes\H_2,\K_1\otimes\K_2)$.

  \begin{Remark}\label{tensorial measure}
    Since $H_0=-\Delta_{\rm D}^\Sigma\otimes1+1\otimes P^2$, the domain of $H_0$ has the following
    form \cite[Sec.~3]{Boutet/Georgescu}:
    \begin{equation*}
      \dom(H_0)=\[\dom(-\Delta^\Sigma_{\rm D})\otimes\ltwo(\R)\]
      \cap\[\ltwo(\Sigma)\otimes\H^2(\R)\].
    \end{equation*}
    The set $\dom(H_0)$ is endowed with the intersection topology, so that it is a Hilbert. The
    spectral measure of $H_0$ admits the tensorial decomposition \cite[Ex.~8.21]{Weidmann80}:
    \begin{equation*}
      E^{H_0}(\:\!\cdot\:\!)
      =\sum_{\alpha\geq1}\P_\alpha\otimes E^{P^2+\nu_\alpha}(\:\!\cdot\:\!)\:\!.
    \end{equation*}
    Hence the equality
    \begin{equation}\label{exponential}
      \e^{itH_0}=\sum_{\alpha\geq1}\P_\alpha\otimes\e^{it(P^2+\nu_\alpha)}
    \end{equation}
    holds in the sense of the strong convergence. Furthermore each $\varphi\in\D^\Omega_t$ is a finite
    sum of vectors $\varphi^\Sigma_\alpha\otimes\varphi^\R_\alpha$, where
    $\varphi^\Sigma_\alpha\in\P_\alpha\ltwo(\Sigma)$ and $\varphi^\R_\alpha\in\D^\R_t$.
  \end{Remark}
  
  For each $r>0$, we define the auxiliary time $\tau_r^{\rm free}(\varphi)$ by
  \begin{align*}
    \tau_r^{\rm free}(\varphi):=\mbox{$\frac12$}\Big\{\int_{-\infty}^0\de t&\Big[
      \left\|F_r\e^{-itH_0}\varphi\right\|^2-\left\|F_r\e^{-itH_0}S\varphi\right\|^2\Big]\\
    &+\int_0^\infty\Big[\left\|F_r\e^{-itH_0}S\varphi\right\|^2
      -\left\|F_r\e^{-itH_0}\varphi\right\|^2\Big]\Big\}\:\!.
  \end{align*}
  The supscript ``free'' makes reference to the fact that the formula for $\tau_r^{\rm free}(\varphi)$
  involves only the free evolution of the vectors $\varphi$ and $S\varphi$.
  
  \begin{Lemma}\label{finite r}
    Suppose that the hypotheses 1 and 2 of Theorem \ref{abstract time delay} hold and let $r>0$,
    $\varphi\in\D^\Omega_0$. Then
    \begin{enumerate}
    \item[(a)] $\left\|F_r\e^{-itH_0}\varphi\right\|$ belongs to $\ltwo(\R,\de t)$,
    \item[(b)] $\left\|F_r\e^{-itH_0}S\varphi\right\|$ belongs to $\ltwo(\R,\de t)$,
    \item[(c)] $\left\|F_r\e^{-itH}W^-\varphi\right\|$ belongs to $\ltwo(\R,\de t)$,
    \item[(d)] $\tau_r(\varphi)$ and $\tau_r^{\rm free}(\varphi)$ exist.
    \end{enumerate}
  \end{Lemma}
  
  \begin{proof}
    Since $F_r=1\otimes\chi_{[-r,r]}(Q)$, the point (a) follows from Remark \ref{tensorial measure}
    and the local smoothness \cite[Thm.~1]{Lavine73} of $\chi_{[-r,r]}(Q)$ with respect to $P^2$.
    Since $S$ and $E^{H_0}(\:\!\cdot\:\!)$ commute, the statement (b) can be shown as (a). The point
    (c) follows from the intertwining relation
    $E^H(\:\!\cdot\:\!)W^\pm=W^\pm E^{H_0}(\:\!\cdot\:\!)$ and the fact that $F_r$ is locally
    $H$-smooth on $(\nu_1,\infty)\setminus\(\sigma_{\rm p}(H)\cup\T\)$. The last statement is a
    consequence of points (a), (b) and (c).
  \end{proof}
  
  The following result can be easily deduced from the proof of \cite[Prop.~2]{Amrein/Cibils}.
  
  \begin{Lemma}\label{same limit}
    Suppose that the hypotheses 1 and 2 of Theorem \ref{abstract time delay} hold and let
    $\varphi\in\D^\Omega_0$ be such that
    \begin{equation*}
      \left\|\(W^--1\)\e^{-itH_0}\varphi\right\|\in\lone((-\infty,0),\de t)
    \end{equation*}
    and
    \begin{equation*}
      \left\|\(W^+-1\)\e^{-itH_0}S\varphi\right\|\in\lone((0,\infty),\de t)\:\!.
    \end{equation*}
    Then one has the equality
    \begin{equation}\label{free delay}
      \lim_{r\to\infty}\tau_r(\varphi)=\lim_{r\to\infty}\tau_r^{\rm free}(\varphi)\:\!.
    \end{equation}
  \end{Lemma}

  We emphasize that the equation (\ref{free delay}) should be interpreted as follows: if one of the
  two limits exists, then so does the other one, and the two limits are equal.
  
  \subsection{Spectral decomposition and trace-type operator}\label{spectral decomposition}
  
  We now gather some results on the spectral transformation for $H_0$ and on the associated 
  trace-type operator. We begin with the definition of the trace-type operator. $\H(\lambda)$
  denotes the fibre at energy $\lambda\geq\nu_1$ for the spectral decomposition of $H_0$:
  \begin{equation*}
    \H(\lambda):=\bigoplus_{\alpha\in\N(\lambda)}
    \left\{\P_\alpha\ltwo(\Sigma)\oplus\P_\alpha\ltwo(\Sigma)\right\},
  \end{equation*}
  where $\N(\lambda):=\{\alpha\in\N\setminus\{0\}\:\!:\:\!\nu_\alpha\leq\lambda\}$. Since
  $\H(\lambda)$ is naturally embedded in
  \begin{equation*}
    \H(\infty):=\bigoplus_{\alpha\geq1}
    \left\{\P_\alpha\ltwo(\Sigma)\oplus\P_\alpha\ltwo(\Sigma)\right\},
  \end{equation*}
  we shall sometimes write $\H(\infty)$ instead of $\H(\lambda)$. For $\xi\in\R$, let
  $\gamma(\xi):\mathscr S(\R)\to\C$ be the trace operator given by
  $\gamma(\xi)\varphi:=\varphi(\xi)$. Then, for $\lambda\in(\nu_1,\infty)\setminus\T$, we define the
  trace-type operator $T(\lambda):\ltwo(\Sigma)\odot\mathscr S(\R)\to\H(\lambda)$ by
  \begin{equation}\label{U1}
    \[T(\lambda)\varphi\]_\alpha:=(\lambda-\nu_\alpha)^{-1/4}
    \left\{\big[\P_\alpha\otimes\gamma(-\sqrt{\lambda-\nu_\alpha})\big]\varphi,
    \big[\P_\alpha\otimes\gamma(\sqrt{\lambda-\nu_\alpha})\big]\varphi\right\}.
  \end{equation}

  In the next lemma we show some regularity properties of the operator $T(\lambda)$. The proof can
  be found in the appendix.

  \begin{Lemma}\label{diff. of trace}
    Let $t\in\R$. Then
    \begin{enumerate}
    \item[(a)] For any $\lambda\in(\nu_1,\infty)\setminus\T$ and $s>1/2$, the operator $T(\lambda)$
      extends to an element of $\B\(\ltwo(\Sigma)\otimes\H^s_t(\R),\H(\infty)\)$.
    \item[(b)] For any $s>1/2$, the function
      $T:(\nu_1,\infty)\setminus\T\to\B\(\ltwo(\Sigma)\otimes\H^s_t(\R),\H(\infty)\)$ is locally
      H\"older continuous.
    \item[(c)] For any $s>n+1/2$, $n\in\N$, the function $\lambda\mapsto T(\lambda)$ is $n$ times
      continuously differentiable as a map from $(\nu_1,\infty)\setminus\T$ to
	$\B\big(\ltwo(\Sigma)\otimes\H^s_t(\R),\H(\infty)\big)$.
    \end{enumerate}
  \end{Lemma}
   
  We give now the spectral transformation for $H_0$ in terms of the operators $T(\lambda)$.
  
  \begin{Proposition}\label{decomposition of H_0}
    The mapping $\U:\H\to\int_{[\nu_1,\infty)}^\oplus\de\lambda\,\H(\lambda)$, defined by
    \begin{equation}\label{U2}
      (\U\varphi)(\lambda):=2^{-1/2}\:\!T(\lambda)(1\otimes\F)\varphi
    \end{equation}
    for all $\varphi\in\ltwo(\Sigma)\odot\mathscr S(\R)$, $\lambda\in(\nu_1,\infty)\setminus\T$,
    is unitary and
    \begin{equation*}
      \U H_0\U^*=\int_{[\nu_1,\infty)}^\oplus\de\lambda\,\lambda\:\!.
    \end{equation*}
  \end{Proposition}

  \begin{proof}
    A direct calculation shows that $\|\U\varphi\|=\|\varphi\|$ for all
    $\varphi\in\ltwo(\Sigma)\odot\mathscr S(\R)$. Since $\ltwo(\Sigma)\odot\mathscr S(\R)$ is dense
    in $\H$, this implies that $\U$ is an isometry. Furthermore, for any
    $\psi\equiv\{\psi^-_\alpha(\lambda),
    \psi^+_\alpha(\lambda)\}\in\int_{[\nu_1,\infty)}^\oplus\de\lambda\,\H(\lambda)$,
    one can check that
    \begin{equation}\label{U3}
      \U^*\psi=\(1\otimes\F^*\)\widetilde\psi\qquad\textrm{where}\qquad
      \widetilde\psi(\:\!\cdot\:\!,\xi):=
      \begin{cases}
	\sqrt{2|\xi|}\sum_{\alpha\geq1}\psi^-_\alpha(\xi^2+\nu_\alpha) & \textrm{if}\quad\xi<0\\
	\sqrt{2|\xi|}\sum_{\alpha\geq1}\psi^+_\alpha(\xi^2+\nu_\alpha) & \textrm{if}\quad\xi\geq0\:\!,
      \end{cases}
    \end{equation}
    so that $\|\U^*\psi\|=\|\psi\|$. Hence $\U$ is unitary. The second statement follows by using
    (\ref{U1}) and (\ref{U2}).
  \end{proof}

  Since the scattering operator $S$ commutes with $H_0$, it follows by Proposition
  \ref{decomposition of H_0} that $S$ admits the direct integral decomposition
  \begin{equation*}
    \U S\U^*=\int_{[\nu_1,\infty)}^\oplus\de\lambda\,S(\lambda)\:\!,
  \end{equation*}
  where $S(\lambda)$ (the $S$-matrix at energy $\lambda$) is an operator acting unitarily in
  $\H(\lambda)$.

  \subsection{Existence theorem}\label{abstract existence}

  In the present section we shall give the proof of Theorem \ref{abstract time delay}. We first prove
  an asymptotic formula involving
  \begin{equation*}
    D_0:=\mbox{$\frac12$}\(P^{-1}Q+QP^{-1}\),
  \end{equation*}
  which is a well defined symmetric operator on $\D^\R_1$.

  \begin{Proposition}\label{general proposition}
    \rule{0.5\textwidth}{0ex}
    \begin{enumerate}
    \item[(a)] Suppose that the hypothesis 2 of Theorem \ref{abstract time delay} holds and let
      $\varphi\in\D^\Omega_0$. Then
      \begin{equation*}
	\tau_r^{\rm free}(\varphi)=\mbox{$\frac12$}\int_0^\infty\de t\,\Big\langle S^*\varphi,
	\Big[1\otimes\Big(\e^{itP^2}\chi_{[-r,r]}(Q)\e^{-itP^2}
	-\e^{-itP^2}\chi_{[-r,r]}(Q)\e^{itP^2}\Big),S\Big]\varphi\Big\rangle\:\!.
      \end{equation*}
    \item[(b)] For all $\varphi,\psi\in\D^\R_2$
      \begin{equation}\label{asymptotic}
	\lim_{r\to\infty}\int_0^\infty\de t\,\big<\varphi,
	\big[\e^{itP^2}\chi_{[-r,r]}(Q)\e^{-itP^2}-\e^{-itP^2}\chi_{[-r,r]}(Q)\e^{itP^2}\big]\psi\big>
	=-\<\varphi,D_0\psi\>.
      \end{equation}
    \item[(c)] Suppose that the hypothesis 2 of Theorem \ref{abstract time delay} holds and let
      $\varphi\in\D^\Omega_2$ be such that $S\varphi\in\D^\Omega_2$. Then
      \begin{equation}\label{Eisenbud 1}
	\lim_{r\to\infty}\tau^{\rm free}_r(\varphi)
	=-\mbox{$\frac12$}\<\varphi,S^*[1\otimes D_0,S]\varphi\>.
      \end{equation}
    \end{enumerate}
  \end{Proposition}
  
  \begin{proof}
    (a) Due to (\ref{exponential}), one has the equality
    \begin{equation*}
      \e^{itH_0}F_r\e^{-itH_0}=1\otimes\e^{itP^2}\chi_{[-r,r]}(Q)\e^{-itP^2}.
    \end{equation*}
    This together with the unitarity of the scattering operator implies the claim.
    
    (b) (i) It is sufficient to prove (\ref{asymptotic}) for $\varphi=\psi$, the case $\varphi\neq\psi$
    being obtained by means of the polarization identity.

    For any $f\in\linf(\R)$ and $t>0$ one has \cite[Eq.~(13.4)]{Amrein/Jauch/Sinha}
    \begin{equation*}
      \e^{itP^2}f(Q)\e^{-itP^2}=Z_{1/4t}^*f(2tP)Z_{1/4t}\:\!,
    \end{equation*}
    where $Z_\tau:=\e^{i\tau Q^2}$. This together with the change of variables $\mu:=r(2t)^{-1}$
    and $\nu:=(2r)^{-1}$ leads to the equality
    \begin{align}
      \int_0^\infty\de t\,\big<\varphi,\big[\e^{itP^2}&\chi_{[-r,r]}(Q)\e^{-itP^2}
      -\e^{-itP^2}\chi_{[-r,r]}(Q)\e^{itP^2}\big]\varphi\big>\nonumber\\
      &=\mbox{$\frac 14$}\int_0^\infty\frac{\de\mu}{\nu\mu^2}\<\varphi,\[Z_{\nu\mu}^*
      \chi_{[-\mu,\mu]}(P)Z_{\nu\mu}-Z_{\nu\mu}\chi_{[-\mu,\mu]}(P)Z_{\nu\mu}^*\]\varphi\>.
      \label{limit0}
    \end{align}
    Hence the l.h.s. of (\ref{asymptotic}) (for $\varphi=\psi$) can be  written as
    \begin{align}\label{limit}
      K_\infty(\varphi):=\lim_{\nu\searrow0}\mbox{$\frac 14$}\int_0^\infty\frac{\de\mu}{\nu\mu^2}\,
      \big\langle\varphi,\big[Z_{\nu\mu}^*&\chi_{[-\mu,\mu]}(P)Z_{\nu\mu}-\chi_{[-\mu,\mu]}(P)\\
	&+\chi_{[-\mu,\mu]}(P)-Z_{\nu\mu}\chi_{[-\mu,\mu]}(P)Z_{\nu\mu}^*\big]\varphi\big\rangle\:\!.
      \nonumber
    \end{align}
    (ii) To prove the statement, we shall show that one may interchange the limit and the integral
    in (\ref{limit}), by invoking the Lebesgue dominated convergence theorem. This will be done in
    (iii) below. If one assumes the result for the moment, then a direct calculation as in
    \cite[Sec.~2]{Amrein/Cibils} leads to the desired equality, that is
    \begin{equation*}
      K_\infty(\varphi)=\mbox{$\frac 14$}\int_0^\infty\frac{\de\mu}{\mu^2}\frac{\de}{\de\nu}
      \<\varphi,\[Z_{\nu\mu}^*\chi_{[-\mu,\mu]}(P)Z_{\nu\mu}
      -Z_{\nu\mu}\chi_{[-\mu,\mu]}(P)Z_{\nu\mu}^*\]\varphi\>\Big|_{\nu=0}=-\<\varphi,D_0\varphi\>
    \end{equation*}
    if $\varphi\in\D^\R_2$.

    (iii) It remains to prove the applicability of the Lebesgue dominated convergence theorem to
    (\ref{limit}). For this we rewrite (\ref{limit0}) (which is equivalent to (\ref{limit})) as
    \begin{align}\label{Lebesgue integrand}
      K_\infty(\varphi)=\lim_{\nu\searrow0}\mbox{$\frac 14$}\int_0^\infty\frac{\de\mu}\mu
      \Big[\Big\langle\chi_{[-\mu,\mu]}(P)&Z_{\nu\mu}\varphi,
	\frac{Z_{\nu\mu}-Z_{\nu\mu}^*}{\nu\mu}\,\varphi\Big\rangle\\
	&+\Big\langle\frac{Z_{\nu\mu}-Z_{\nu\mu}^*}{\nu\mu}\,\varphi,
	\chi_{[-\mu,\mu]}(P)Z_{\nu\mu}^*\varphi\Big\rangle\Big]\:\!.\nonumber
    \end{align}
    Since $\tau^{-1}(Z_\tau-Z_\tau^*)\varphi$ converges strongly to $2iQ^2\varphi$ as $\tau\to0$,
    we may choose a number $\delta>0$ such that
    $\|\tau^{-1}(Z_\tau-Z_\tau^*)\varphi\|\leq3\|Q^2\varphi\|$ for all $\tau\in[-\delta,\delta]$. We
    then have
    \begin{equation}\label{majoration1}
      \left\|\mbox{$\frac1{\nu\mu}$}(Z_{\nu\mu}-Z_{\nu\mu}^*)\varphi\right\|\leq
      \begin{cases}
	3\|Q^2\varphi\| & \textrm{if}\quad\nu\mu\leq\delta\\
	\frac2\delta\|\varphi\| & \textrm{if}\quad\nu\mu\geq\delta\:\!.
      \end{cases}
    \end{equation}
    Let $\ell\in(0,1/2)$, then $|P|^{-\ell}\<Q\>^{-2}$ belongs to $\B(\ltwo(\R))$ (after
    exchanging the role of $P$ and $Q$, this follows from the fact that $|Q|^{-\ell}$ is
    $P^2$-bounded \cite[Prop.~2.28]{Amrein81}), and
    \begin{equation*}
      |\mu^{-1}\xi|^{\ell}\chi_{[-\mu,\mu]}(\xi)\leq\chi_{[-\mu,\mu]}(\xi)\leq1
    \end{equation*}
    for all $\xi\in\R$. Thus one has the estimate
    \begin{align}\label{majoration2}
      \mu^{-1}\left\|\chi_{[-\mu,\mu]}(P)Z_{\pm\nu\mu}\varphi\right\|\nonumber
      &=\mu^{\ell-1}\big\||\mu^{-1}P|^{\ell}\chi_{[-\mu,\mu]}(P)
      |P|^{-\ell}\<Q\>^{-2}Z_{\pm\nu\mu}\<Q\>^2\varphi\big\|\nonumber\\
      &\leq\textrm{Const.}\;\mu^{\ell-1}\big\|\<Q\>^2\varphi\big\|.
    \end{align}
    Hence (\ref{majoration1}) and (\ref{majoration2}) imply that the integrand in
    (\ref{Lebesgue integrand}) is bounded by a function in $\loneloc((0,\infty),\de\mu)$, which is
    sufficient for applying the Lebesgue dominated convergence theorem on any finite interval
    $[0,\mu_0]$.

    Since the case $\mu\to\infty$ can be treated as in \cite[Sec.~2]{Amrein/Cibils}, this concludes
    the proof of the statement.
    
    (c) This is a consequence of Remark \ref{tensorial measure} and points (a) and (b).
  \end{proof}

  \begin{Remark}\label{Eisenbud remark}
    We know from Section \ref{spectral decomposition} that $\H$ can be identified with the direct
    integral $\int_{[\nu_1,\infty)}^\oplus\de\lambda\,\H(\lambda)$, where $H_0$ acts as the
    multiplication operator by $\lambda$. So one may write $\varphi(\lambda)$ for the component of
    $\varphi\in\H$ at energy $\lambda$ and $\<\:\!\cdot\:\!,\:\!\cdot\:\!\>_{\H(\lambda)}$ for the
    scalar product in $\H(\lambda)$. A direct calculation using (\ref{U1})--(\ref{U3}) shows that
    $1\otimes D_0=2i\:\!\frac\de{\de\lambda}$ in the spectral representation of $H_0$. On the other
    hand $\varphi\in\dom(1\otimes D_0^2)$ if $\varphi\in\D^\Omega_2$. Therefore if
    $\varphi\in\D^\Omega_2$, then the function $\lambda\mapsto\varphi(\lambda)$ is continuously
    differentiable on each interval $(\nu_\alpha,\nu_{\alpha+1})$. As a consequence, if
    $\varphi\in\D^\Omega_2$ is such that $S\varphi\in\D^\Omega_2$, and if the function
    $\lambda\mapsto S(\lambda)$ is strongly continuously differentiable on the support of
    $\varphi(\:\!\cdot\:\!)$, then one gets from (\ref{Eisenbud 1}) the equalities
    \begin{equation}\label{Eisenbud}
      \lim_{r\to\infty}\tau^{\rm free}_r(\varphi)=-i\int_{\nu_1}^\infty\de\lambda\,
      \Big\langle\varphi(\lambda),S(\lambda)^*\Big[\frac{\de S(\lambda)}{\de\lambda}\Big]
      \varphi(\lambda)\Big\rangle_{\H(\lambda)}
      \equiv\<\varphi,\tau_\textsc{e-w}\varphi\>.
    \end{equation}
    Provided that (\ref{free delay}) holds, (\ref{Eisenbud}) expresses
    the identity of the (global) time delay and the Eisenbud-Wigner time delay in waveguides.
  \end{Remark}

  Theorem \ref{abstract time delay} is a direct consequence of Lemma \ref{finite r}, Lemma
  \ref{same limit}, Proposition \ref{general proposition} and Remark \ref{Eisenbud remark}.

  \begin{Remark}\label{multichannel}
    The $S$-matrix at energy $\lambda$ can be written as the double sum
    \begin{equation*}
      S(\lambda)=\sum_{\beta,\alpha\in\N(\lambda)}S_{\beta\alpha}(\lambda)\:\!,
    \end{equation*}
    where $S_{\beta\alpha}(\lambda):=\[\U(\P_\beta\otimes1)S(\P_\alpha\otimes1)\U^*\](\lambda)$.
    Therefore if $\varphi_\alpha$ is a vector in $(\P_\alpha\otimes1)\H$ satisfying the hypotheses
    of Theorem \ref{abstract time delay}, then a simple calculation shows that (\ref{Eisenbud}) is
    equivalent to
    \begin{equation}\label{Eisenbud 2}
      \lim_{r\to\infty}\tau^{\rm free}(\varphi_\alpha)=-i
      \int_{\nu_1}^\infty\de\lambda\,\bigg\langle\varphi_\alpha(\lambda),\sum_{\beta\in\N(\lambda)}
      S_{\beta\alpha}(\lambda)^*\Big[\frac{\de S_{\beta\alpha}(\lambda)}{\de\lambda}\Big]
      \varphi_\alpha(\lambda)\bigg\rangle_{\H(\lambda)}\:\!.
    \end{equation}
    This equation admits a natural interpretation: if each subspace $(\P_\alpha\otimes1)\H$ is seen
    as a channel Hilbert space, then (\ref{Eisenbud 2}) can be considered as a multichannel
    formulation in waveguides of the identity of the (global) time delay and the Eisenbud-Wigner
    time delay for an incoming state in channel $\alpha$.
  \end{Remark}
  
  \section{Time delay in waveguides: the short-range case}
  \setcounter{equation}{0}

  \subsection{Short-range scattering in waveguides}\label{short-range scattering}

  In this section we collect some results on the scattering theory for the pair $\{H_0,H\}$ in the
  case $H:=H_0+V$, where $V$ is a short-range potential satisfying the following condition:
  
  \begin{Assumption}\label{assumption V}
    $V$ is a multiplication operator by a real-valued measurable function on $\Omega$ such that $V$
    defines a compact operator from $\dom(H_0)$ to $\H$ and a bounded operator from
    $\ltwo(\Sigma)\otimes\H^2(\R)$ to $\ltwo(\Sigma)\otimes\H_\kappa(\R)$ for some $\kappa>1$.
  \end{Assumption}

  By using duality, interpolation and the fact that $V$ commutes with the operator $1\otimes\<Q\>^t$,
  $t\in\R$, one shows that $V$ also defines a bounded operator from
  $\ltwo(\Sigma)\otimes\H^{2s}_t(\R)$ to $\ltwo(\Sigma)\otimes\H^{2(s-1)}_{t+\kappa}(\R)$ for any
  $s\in[0,1]$, $t\in\R$.
  
  If $V$ satisfies Assumption \ref{assumption V}, then the operator $H$ is selfadjoint on
  $\dom(H)=\dom(H_0)$, $(H+i)^{-1}-(H_0+i)^{-1}$ is compact and
  $\sigma_{\rm ess}(H)=\sigma_{\rm ess}(H_0)=[\nu_1,\infty)$. In order to get more informations on
  $H$, we shall apply the conjugate operator method. We refer to \cite{ABG} for the definitions of
  the regularity classes appearing in the sequel, and for more explanations on the conjugate
  operator method.

  For $\varepsilon\in(0,1)$, we choose a function $\vartheta\in C^\infty_0\((\varepsilon,\infty)\)$
  and define $F:\R\to\R$ by
  \begin{equation*}
    F(x):=
    \begin{cases}
      \frac1{2x}\:\!\vartheta(x^2) & \textrm{if}\quad x\in(-\infty,-\sqrt\varepsilon)
      \cup(\sqrt\varepsilon,\infty)\\
      0 & \textrm{otherwise}.
    \end{cases}
  \end{equation*}
  We first introduce the operator $A_\shortparallel:=F(P)Q+\mbox{$\frac i2$}F'(P)$ acting on
  $\mathscr S(\R)$. $A_\shortparallel$ has the following properties \cite[Lemma~7.6.4]{ABG}:
  $A_\shortparallel$ is essentially selfadjoint, the group
  $\{\e^{i\tau A_{\shortparallel}}\}_{\tau\in\R}$ leaves $\dom(-\Delta^\R)=\H^2(\R)$ invariant,
  $-\Delta^\R$ is of class $C^\infty(A_\shortparallel)$ and $A_\shortparallel$ is strictly
  conjugate to $-\Delta^\R$ on $(-\infty,0)\cup I_\vartheta$, where
  $I_\vartheta:=\{u\in(\varepsilon,\infty)\:\!:\:\!\vartheta(u)=1\}$. Now let
  $A:=1\otimes A_\shortparallel$. It turns out that $H_0$ has many regularity properties with
  respect to $A$, namely (see \cite[Sec.~3]{Boutet/Georgescu}) $\{\e^{i\tau A}\}_{\tau\in\R}$ is a
  $C_0$-group in $\dom(H_0)$, $H_0$ is of class $C^\infty(A)$ and $A$ is strictly conjugate to
  $H_0$ on $(-\infty,\nu_1)\cup J_\vartheta$, where $J_\vartheta$ is a bounded open set in
  $(\nu_1,\infty)\setminus\T$ depending on $I_\vartheta$. The exact nature of $J_\vartheta$ can be
  explicitly deduced from that of $I_\vartheta$ by using the formula
  \cite[Eq.~(3.8)]{Boutet/Georgescu}, which relates the Mourre estimate for $-\Delta^\R$ to the
  Mourre estimate for $H_0$. In our case it is enough to note that, given any compact set $K$ in
  $\R\setminus\T$, there exist $\varepsilon\in(0,1)$ and
  $\vartheta\in C^\infty_0\((\varepsilon,\infty)\)$ such that $K$ is contained in
  $(-\infty,\nu_1)\cup J_\vartheta$.

  Now we prove that $V$ also satisfies regularity conditions with respect to $A$. Given an
  operator $B$ in $\H$ and a Hilbert space $\G\subset\H$, we write
  $\dom(B;\G):=\{\varphi\in\dom(B)\cap\G\:\!:\:\!B\varphi\in\G\}$ for the domain of $B$ in $\G$.

  \begin{Lemma}\label{V regularity}
    Let $\:\!V$ satisfy Assumption \ref{assumption V}. Then
    \begin{enumerate}
    \item[(a)] $V$ is of class $\mathscr C^{1,1}(A;\dom(H_0),\dom(H_0)^*)$.
    \item[(b)] The operators $\[H_0,A\]$ and $\[H,A\]$, which a priori only belong to
      $\B\(\dom(H_0),\dom(H_0)^*\)$, are such that $\[H_0,A\]\in\B(\dom(H_0))$ and
      $\[H,A\]\in\B(\dom(H_0),\H)$.
    \end{enumerate}
  \end{Lemma}

  \begin{proof}
    (a) We use the criterion \cite[Thm.~7.5.8]{ABG} to prove the statement. The three conditions
    needed for that theorem are obtained in points (i), (ii) and (iii) below.

    (i) Let $\Lambda:=1\otimes\<Q\>$. Since $\{\e^{i\tau\<Q\>}\}_{\tau\in\R}$ is a polynomially
    bounded $C_0$-group in $\H^2(\R)$ \cite[Sec.~7.6.3]{ABG}, a direct calculation using the
    tensorial decomposition of $H_0$ (see Remark \ref{tensorial measure}) shows that
    $\{\e^{i\tau\Lambda}\}_{\tau\in\R}$ is a polynomially bounded $C_0$-group in $\dom(H_0)$.

    (ii) Since $\{\e^{i\tau A}\}_{\tau\in\R}$ is a $C_0$-group in $\dom(H_0)$, there exists $r>0$
    such that $-ir$ belongs to the resolvent set of $A$ (considered as an operator in $\dom(H_0)$).
    In particular, the operator $(A+ir)^{-1}=-i\int_0^\infty\de\tau\,\e^{-r\tau}\e^{i\tau A}$ is a
    homeomorphism from $\dom(H_0)$ onto $\dom\(A;\dom(H_0)\)$ (both domains being endowed with their
    natural graph topology). Therefore any set $\E$ of the form $(A+ir)^{-1}\D$, with $\D$ dense in
    $\dom(H_0)$, is dense in $\dom\(A;\dom(H_0)\)$. Let us take
    $\D:=\{\varphi_\alpha\}\odot\mathscr S(\R)$, where $\{\varphi_\alpha\}$ is the set of
    eigenvectors of $-\Delta^\Sigma_{\rm D}$ (since $H_0\upharpoonright\D$ is essentially
    selfadjoint, $\D$ is dense in $\dom(H_0)$). A vector $\psi$ in $\E$ is of the form
    $
    \psi=-i\sum_{\alpha\leq\mathrm{Const.}}\varphi_\alpha\otimes
    \int_0^\infty\de\tau\,\e^{-r\tau}\e^{i\tau A_\shortparallel}\eta_\alpha
    $,
    where $(\varphi_\alpha,\eta_\alpha)\in\{\varphi_\alpha\}\times\mathscr S(\R)$ and the integral
    converges in $\H^2(\R)$. Since $\<Q\>^{-2}\in\B\(\ltwo(\R)\)$ and
    $A_\shortparallel\eta_\alpha\in\mathscr S(\R)$, the vector
    \begin{equation*}
      \widetilde\psi:=-i\sum_{\alpha\leq\mathrm{Const.}}\varphi_\alpha\otimes\int_0^\infty\de\tau\,
      \e^{-r\tau}\<Q\>^{-2}\e^{i\tau A_\shortparallel}A_\shortparallel\eta_\alpha
    \end{equation*}
    belongs to $\H$. Furthermore $\widetilde\psi=\Lambda^{-2}A\psi$ and
    $\Lambda^{-2}A\psi\in\dom(H_0)$. Since $\e^{i\tau A_\shortparallel}\eta_\alpha\in\mathscr S(\R)$
    \cite[Prop.~4.2.4]{ABG}, one can use commutator expansions to get the equality
    \begin{equation*}
      \big\|\<Q\>^{-2}\e^{i\tau A_\shortparallel}A_\shortparallel\eta_\alpha
      -S_1\<Q\>^{-1}\e^{i\tau A_\shortparallel}\eta_\alpha\big\|_{\H^2(\R)}=0
    \end{equation*}
    for some operator $S_1\in\B\(\H^2(\R)\)$. This implies that
    \begin{equation}\label{equality in dom(H_0)}
      \big\|\Lambda^{-2}A\psi-(1\otimes S_1)\Lambda^{-1}\psi\big\|_{\dom(H_0)}=0
    \end{equation}
    for $\psi\in\E$. Since $1\otimes S_1$ and $\Lambda^{-1}$ belong to $\B\(\dom(H_0)\)$ and $\E$
    is dense in $\dom\(A;\dom(H_0)\)$, (\ref{equality in dom(H_0)}) even holds for
    $\psi\in\dom\(A;\dom(H_0)\)$. Thus, for each $\psi\in\dom\(A^2;\dom(H_0)\)$, one gets
    \begin{equation*}
      \left\|\Lambda^{-2}A^2\psi-(1\otimes S_1)\Lambda^{-1}A\psi\right\|_{\dom(H_0)}
      =\left\|(\Lambda^{-2}A)A\psi-(1\otimes S_1)\Lambda^{-1}A\psi\right\|_{\dom(H_0)}=0\:\!.
    \end{equation*}
    Using an argument similar to the one leading to (\ref{equality in dom(H_0)}), one shows that
    
    \begin{equation*}
      \big\|\Lambda^{-1}A\psi-(1\otimes S_2)\psi\big\|_{\dom(H_0)}=0
    \end{equation*}
    for each $\psi\in\dom\(A;\dom(H_0)\)$ and some operator $S_2\in\B\(\H^2(\R)\)$. Therefore
    \begin{equation*}
      \left\|\Lambda^{-2}A^2\psi-(1\otimes S_1S_2)\psi\right\|_{\dom(H_0)}=0
    \end{equation*}
    for each $\psi\in\dom\(A^2;\dom(H_0)\)$. This implies that
    $\Lambda^{-2}A^2:\dom\(A^2;\dom(H_0)\)\to\dom(H_0)$ extends to an element of $\B\(\dom(H_0)\)$.

    (iii) The short-range decay of $V$ required in \cite[Eq.~(7.5.29)]{ABG} follows from Assumption
    \ref{assumption V}.
    
    (b) We have $\[H_0,A\]\in\B(\dom(H_0))$ because $\[H_0,iA\]=1\otimes\vartheta(P^2)$
    \cite[Lemma~7.6.4]{ABG}, \cite[Sec.~3]{Boutet/Georgescu}. Since $H=H_0+V$, it remains to show
    that $\[V,A\]\in\B(\dom(H_0),\H)$. This follows by using the fact that $V$ is bounded from
    $\ltwo(\Sigma)\otimes\H^{2s}_t(\R)$ to $\ltwo(\Sigma)\otimes\H^{2(s-1)}_{t+\kappa}(\R)$ for any
    $s\in[0,1]$, $t\in\R$, and the fact that $A$ is bounded from $\ltwo(\Sigma)\otimes\H^s_t(\R)$
    to $\ltwo(\Sigma)\otimes\H^s_{t-1}(\R)$ for any $s,t\in\R$.
  \end{proof}

  Since $\{\e^{i\tau A}\}_{\tau\in\R}$ leaves $\dom(H_0)$ invariant and $H_0$ is of class $C^\infty(A)$,
  Lemma \ref{V regularity}.(a) implies that $H$ is of class $\mathscr C^{1,1}(A)$
  \cite[Thm.~6.3.4.(b)]{ABG}. This has the following consequence.
  
  \begin{Lemma}\label{Mourre for H}
    Let $\:\!V$ satisfy Assumption \ref{assumption V}. Then $A$ is conjugate to $H$ on
    $(-\infty,\nu_1)\cup J_\vartheta$.
  \end{Lemma}

  \begin{proof}
    Since $H_0$ and $H$ are of class $\mathscr C^{1,1}(A)$, $(H+i)^{-1}-(H_0+i)^{-1}$ is compact and
    $A$ is strictly conjugate to $H_0$ on $(-\infty,\nu_1)\cup J_\vartheta$, the claim follows by
    \cite[Thm.~7.2.9]{ABG}. 
  \end{proof}

  Now we can prove limiting absorption principles for $H_0$ and $H$, and state spectral properties
  of $H$. If $\G^\mu:=\dom(H_0^\mu)$, $\mu\in\R$, then the limiting absorption
  principles can be expressed in terms of the Banach space
  $\K:=\(\G^{-1/2}\cap\dom(A;\G^{-1}),\G^{-1/2}\)_{1/2,1}$ defined by real interpolation
  \cite[Chap.~2]{ABG}. We emphasize that $\K$ contains $\ltwo(\Sigma)\otimes\H^{-1}_t(\R)$ for any
  $t>1/2$, which is shown in the appendix.

  \begin{Theorem}\label{limiting absorption}
    Let $\:\!V$ satisfy Assumption \ref{assumption V}. Then
    \begin{enumerate}
    \item[(a)] $H$ has no singularly continuous spectrum.
    \item[(b)] The eigenvalues of $H$ in $\sigma(H)\setminus\T$ are of finite multiplicity and can
      accumulate at points of $\T$ only.
    \item[(c)] The limit $\lim_{\varepsilon\searrow0}(H_0-\lambda\mp i\varepsilon)^{-1}$,
      resp. $\lim_{\varepsilon\searrow0}(H-\lambda\mp i\varepsilon)^{-1}$, exists in the
      weak$^*$ topology of $\B(\K,\K^*)$ uniformly in $\lambda$ on each compact subset of
      $\:\!\R\setminus\T$, resp. $\R\setminus\(\sigma_{\rm p}(H)\cup\T\)$.
    \end{enumerate}
  \end{Theorem}

  \begin{proof}
    The operator $H$ is of class $\mathscr C^{1,1}(A)$ and $A$ is conjugate to $H$ on
    $(-\infty,\nu_1)\cup J_\vartheta$ by Lemma \ref{Mourre for H}. Furtheremore, given any compact
    set $K$ in $\R\setminus\T$, there exist $\varepsilon\in(0,1)$ and
    $\vartheta\in C^\infty_0\((\varepsilon,\infty)\)$ such that $K$ is contained in
    $(-\infty,\nu_1)\cup J_\vartheta$. Therefore the assertions (a) and (b) follow by the conjugate
    operator method \cite[Cor.~7.2.11\,\&\,Thm.~7.4.2]{ABG}. Due to Lemma \ref{V regularity}.(b) and
    the regularity properties of $H_0$ and $H$ with respect to $A$, the limiting absorption
    principles are obtained via \cite[Thm.~7.5.2]{ABG}.
  \end{proof}

  \begin{Corollary}\label{wave operators}
    Let $\:\!V$ satisfy Assumption \ref{assumption V}. Then
    \begin{enumerate}
    \item[(a)] If $T$ belongs to $\B\(\ltwo(\Sigma)\otimes\H^1_{-t}(\R),\H\)$ for some $t>1/2$, then
      $T$ is locally $H_0$-smooth (resp. $H$-smooth) on $\R\setminus\T$ (resp.
      $\R\setminus\(\sigma_{\rm p}(H)\cup\T\)$).
    \item[(b)] The wave operators $W^\pm$ exist and are complete.
    \end{enumerate}
  \end{Corollary}
  
  \begin{proof}
    (a) Let $\mathscr E:=\ltwo(\Sigma)\otimes\H^{-1}_t(\R)$. Since $\mathscr E\subset\dom(H_0)^*$
    densely, and $\mathscr E\subset\K$, it is enough to verify the remaining hypothesis of
    \cite[Prop.~7.1.3.(b)]{ABG} on $\mathscr E$ to prove the statement. Let $\mathscr E^{*\circ}$ be
    the closure of $\dom(H_0)$ in $\mathscr E^*$, equipped with the norm of $\mathscr E^*$.
    Clearly $\mathscr E^{*\circ}\subset\mathscr E^*$. Furthermore, since $\dom(H_0)$ is
    dense in $\mathscr E^*$, we also have $\mathscr E^*\subset\mathscr E^{*\circ}$. Therefore
    $\mathscr E^*=\mathscr E^{*\circ}$. By taking the adjoint, this leads to
    $\mathscr E=\(\mathscr E^{*\circ}\)^*$.
    
    (b) By the point (a), $V_1:=1\otimes\<Q\>^{-\kappa/2}\<P\>$ is locally $H_0$-smooth on
    $\R\setminus\T$ and $V_2:=(1\otimes\<Q\>^{\kappa/2}\<P\>^{-1})V$ is locally $H$-smooth on
    $\R\setminus\(\sigma_{\rm p}(H)\cup\T\)$. Since $\sigma_{\rm p}(H)\cup\T$ is countable and
    $\<\varphi,V\psi\>=\<V_1\varphi,V_2\psi\>$ for all $\varphi,\psi\in\dom(H_0)$, one can
    conclude by applying the smooth perturbation theory
    \cite[Corollary~to~Thm.~XIII.31]{Reed/Simon}.
  \end{proof}

  Under Assumption \ref{assumption V} one could also find optimal spaces where the analogue of
  the limiting absorption principles of Theorem \ref{limiting absorption}.(c) holds in norm. The
  following particular result is sufficient for us. If $t>1/2$, then the boundary values
  \begin{equation*}
    R^{H_0}(\lambda\pm i0):=\lim_{\varepsilon\searrow0}(H_0-\lambda\mp i\varepsilon)^{-1},\quad
    \lambda\in\R\setminus\T,
  \end{equation*}
  and
  \begin{equation*}
    R^H(\lambda\pm i0):=\lim_{\varepsilon\searrow0}(H-\lambda\mp i\varepsilon)^{-1},\quad
    \lambda\in\R\setminus\(\sigma_{\rm p}(H)\cup\T\),
  \end{equation*}
  exist in $\B\big(\ltwo(\Sigma)\otimes\H_t(\R),\ltwo(\Sigma)\otimes\H_{-t}(\R)\big)$
  (see \cite[Thm.~4.13]{BGM}). In the rest of the section we study the norm differentiability of
  the function $\lambda\mapsto S(\lambda)$, which relies on the differentiability of the function
  $\lambda\mapsto R^H(\lambda\pm i0)$.

  \begin{Lemma}\label{diff. of R}
    Let $t>n+1/2$, $n\in\N$. Let $V$ satisfy Assumption \ref{assumption V} with $\kappa>n+1$. Then
    $\lambda\mapsto R^H(\lambda+i0)$ is $n$ times continuously differentiable as a map from
    $(\nu_1,\infty)\setminus\(\sigma_{\rm p}(H)\cup\T\)$ to
    $\B\big(\ltwo(\Sigma)\otimes\H_t(\R),\ltwo(\Sigma)\otimes\H_{-t}(\R)\big)$.
  \end{Lemma}

  \begin{proof}
    Since $H_0$ is of class $C^\infty(A)$ and $\ltwo(\Sigma)\otimes\H_t(\R)\subset\dom(\<A\>^t)$,
    we have the following result \cite[Sec.~1.7]{BGS}. For each
    $\lambda\in(\nu_1,\infty)\setminus\T$ and $k\leq n$, the boundary values
    $\lim_{\varepsilon\searrow0}(H_0-\lambda\mp i\varepsilon)^{-k-1}$ exist in
    $\B\big(\ltwo(\Sigma)\otimes\H_t(\R),\ltwo(\Sigma)\otimes\H_{-t}(\R)\big)$. Furthermore
    $\lambda\mapsto R^{H_0}(\lambda\pm i0)$ is $k$ times continuously differentiable as a map from
    $(\nu_1,\infty)\setminus\T$ to
    $\B\big(\ltwo(\Sigma)\otimes\H_t(\R),\ltwo(\Sigma)\otimes\H_{-t}(\R)\big)$ with
    \begin{equation*}
      \mbox{$\frac{\de^k}{\de\lambda^k}$}\:\!R^{H_0}(\lambda\pm i0)
      =k!\lim_{\varepsilon\searrow0}(H_0-\lambda\mp i\varepsilon)^{-k-1}.
    \end{equation*}
    Thus one can apply the inductive method of \cite[Lemma~4.3]{Jensen/Nakamura} to infer the
    result for $H$ from the one for $H_0$.
  \end{proof}

  In the following lemma we prove the usual formula for the $S$-matrix.
  
  \begin{Lemma}\label{S formula}
    Let $V$ satisfy Assumption \ref{assumption V}. Then for each
    $\lambda\in(\nu_1,\infty)\setminus\(\sigma_{\rm p}(H)\cup\T\)$, one has the equality
    \begin{equation}\label{S-matrix}
      S(\lambda)=1-i\pi\:\!T(\lambda)\(1\otimes\F\)\[1-VR^H(\lambda+i0)\]
      V\(1\otimes\F^*\)T(\lambda)^*.
    \end{equation}
  \end{Lemma}

  \begin{proof}
    The claim is a consequence of the stationary method \cite[Thm.~6.3]{Kuroda73} applied to the pair
    $\{H_0,H\}$. Therefore we simply verify the principal hypotheses of that theorem.

    The total Hamiltonian admits the factorization $H=H_0+V_1V_2$ where $V_1$ is the $H_0$-compact 
    operator $1\otimes\<Q\>^{-\kappa/2}$ (see \cite[Lemma~2.1]{Krejcirik/Tiedra}) and $V_2$ is the
    (maximal) operator associated to $1\otimes\<Q\>^{\kappa/2}V$. Moreover, since
    $T:(\nu_1,\infty)\setminus\T\to\B\(\ltwo(\Sigma)\otimes\H^s_t(\R),\H(\infty)\)$ is locally
    H\"older continuous for each $t\in\R$, $s>1/2$, the functions
    $T(\:\!\cdot\:\!;V_j):(\nu_1,\infty)\setminus\T\to\B\(\H,\H(\infty)\)$, $j=1,2$, defined by
    \begin{equation*}
      T(\lambda;V_j)\varphi:=\(\U V_j^*\varphi\)(\lambda)\:\!,
    \end{equation*}
    are locally H\"older continuous.
  \end{proof}

  Finally we have the following result on the norm differentiability of the function
  $\lambda\mapsto S(\lambda)$.

  \begin{Proposition}\label{S diff}
    Let $V$ satisfy Assumption \ref{assumption V} with $\kappa>n+1$, $n\in\N$. Then
    $\lambda\mapsto S(\lambda)$ is $n$ times continuously differentiable as a map from
    $(\nu_1,\infty)\setminus\(\sigma_{\rm p}(H)\cup\T\)$ to $\H(\infty)$.
  \end{Proposition}

  \begin{proof}
    Due to (\ref{S-matrix}) and Lemmas \ref{diff. of trace}.(c) and \ref{diff. of R}, all operators
    in the expression for $S(\lambda)$ are $n$ times continuously norm differentiable. Then a direct
    calculation as in the proof of \cite[Thm.~3.5]{Jensen81} implies the claim.
  \end{proof}

  \subsection{Existence theorem}\label{short-range existence}

  To illustrate Theorem \ref{abstract time delay}, we verify in this section the existence of the
  (global) time delay in the case $H:=H_0+V$, where $V$ satisfies Assumption \ref{assumption V} with
  $\kappa>4$. To begin with we prove two technical lemmas in relation with the hypotheses of Theorem
  \ref{abstract time delay}.

  \begin{Lemma}\label{l1 integrals}
    If $\:\!V$ satisfies Assumption \ref{assumption V} with $\kappa>2$ and $\varphi\in\D^\Omega_\tau$
    for some $\tau>2$, then
    \begin{equation}
      \left\|\(W^--1\)\e^{-itH_0}\varphi\right\|\in\lone((-\infty,0),\de t)\label{l1-}
    \end{equation}
    and
    \begin{equation}
      \left\|\(W^+-1\)\e^{-itH_0}\varphi\right\|\in\lone((0,\infty),\de t)\:\!.\label{l1+}
    \end{equation}
  \end{Lemma}

  \begin{proof}
    For $\varphi\in\D^\Omega_\tau$ and $t\in\R$, we have (see the proof of
    \cite[Lemma~4.6]{Jensen81})
    \begin{equation*}
      \(W^--1\)\e^{-itH_0}\varphi=-i\e^{-itH}\int_{-\infty}^t\de s\,\e^{isH}V\e^{-isH_0}\varphi\:\!,
    \end{equation*}
    where the integral is strongly convergent. Hence to prove (\ref{l1-}) it is enough to show that
    \begin{equation}
      \int_{-\infty}^{-\delta}\de t\int_{-\infty}^t\de s\left\|V\e^{-isH_0}\varphi\right\|<\infty
      \label{l1 condition}
    \end{equation}
    for some $\delta>0$. We know from Remark \ref{tensorial measure} that
    $\varphi=\sum_{\alpha\leq\textrm{Const.}}\varphi_\alpha^\Sigma\otimes\varphi_\alpha^\R$, where
    $\varphi_\alpha^\Sigma\in\P_\alpha\ltwo(\Sigma)$ and $\varphi_\alpha^\R\in\D_\tau^\R$. Thus
    there exists $\eta\in C^\infty_0((0,\infty))$ such that $1\otimes\eta(P^2)\varphi=\varphi$.
    Furthermore, if $\zeta:=\min\{\kappa,\tau\}$, then
    $\big\|\<Q\>^\zeta\varphi_\alpha^\R\big\|<\infty$ and $V(1\otimes\<P\>^{-2}\<Q\>^\zeta)$ belongs
    to $\B(\H)$ due to Assumption \ref{assumption V}. This implies that
    \begin{align*}
      \left\|V\e^{-isH_0}\varphi\right\|&\leq\sum_{\alpha\leq\textrm{Const.}}
      \big\|V(1\otimes\<P\>^{-2}\<Q\>^\zeta)\big[\varphi_\alpha^\Sigma\otimes\<Q\>^{-\zeta}\<P\>^2
      \eta(P^2)\e^{-isP^2}\<Q\>^{-\zeta}\<Q\>^\zeta\varphi_\alpha^\R\big]\big\|\\
      &\leq\textrm{Const.}\,\big\|\<Q\>^{-\zeta}\<P\>^2\eta(P^2)\e^{-isP^2}\<Q\>^{-\zeta}\big\|\:\!.
    \end{align*}
    For each $\varepsilon>0$, it follows from \cite[Lemma~9]{ACS} that there exists a constant
    $\textsc c>0$ such that
    $\left\|V\e^{-isH_0}\varphi\right\|\leq\textsc c\(1+|s|\)^{-\zeta+\varepsilon}$. Since
    $\zeta>2$, this implies (\ref{l1 condition}). The proof of (\ref{l1+}) is similar.
  \end{proof}

  Let $\E$ be the finite span of vectors $\varphi\in\H$ of the form
  $\{\varphi(\lambda)\}=\{\rho(\lambda)h(\lambda)\}$ in the spectral representation of $H_0$, where
  $\rho:(\nu_1,\infty)\to\C$ is three times continuously differentiable and has compact support in
  $(\nu_1,\infty)\setminus\(\sigma_{\rm p}(H)\cup\T\)$, and $\lambda\mapsto h(\lambda)\in\H(\lambda)$
  is $\lambda$-independent on each interval $(\nu_\alpha,\nu_{\alpha+1})$. Clearly the set $\E$ is
  dense in $\H$. Furthermore one has the following inclusions.

  \begin{Lemma}\label{inclusions}
    \rule{0.5\textwidth}{0ex}
    \begin{enumerate}
      \item[(a)] $\E$ is contained in $\D^\Omega_3$.
      \item[(b)] Let $\:\!V$ satisfy Assumption \ref{assumption V} with $\kappa>4$. Then $S\E$ is
	contained in $\D^\Omega_3$.
    \end{enumerate}
  \end{Lemma}

  \begin{proof} 
    (a) Let $\varphi\in\E$. It is clear that there exists a compact set $J$ in
    $(\nu_1,\infty)\setminus\(\sigma_{\rm p}(H)\cup\T\)$ such that $E^{H_0}(J)\varphi=\varphi$. Thus,
    in order to show that $\varphi\in\D^\Omega_3$, one has to verify that
    $\varphi\in\ltwo(\Sigma)\otimes\H_3(\R)=\dom(1\otimes Q^3)$.

    Let $\psi\in\ltwo(\Sigma)\odot\mathscr S(\R)$. Then, using (\ref{U1})--(\ref{U3}), we obtain
    \begin{equation}\label{Q3}
      \[\U(1\otimes Q^3)\psi\]_\alpha(\lambda)=\{ig^-_\alpha(\lambda),-ig^+_\alpha(\lambda)\}\:\!,
    \end{equation}
    where
    \begin{align}
      g^\pm_\alpha(\lambda)
      :=\mbox{$\frac38$}(\lambda-\nu_\alpha)^{-3/2}&(\U\psi)^\pm_\alpha(\lambda)
      +\mbox{$\frac32$}(\lambda-\nu_\alpha)^{-1/2}
      \mbox{$\frac\de{\de\lambda}$}(\U\psi)^\pm_\alpha(\lambda)\label{Q3bis}\\
      &+18(\lambda-\nu_\alpha)^{1/2}
      \mbox{$\frac{\de^2}{\de\lambda^2}$}(\U\psi)^\pm_\alpha(\lambda)
      +8(\lambda-\nu_\alpha)^{3/2}
      \mbox{$\frac{\de^3}{\de\lambda^3}$}(\U\psi)^\pm_\alpha(\lambda)\:\!.\nonumber
    \end{align}
    
    The r.h.s. of (\ref{Q3})--(\ref{Q3bis}) with $\psi\in\ltwo(\Sigma)\odot\mathscr S(\R)$ replaced
    by $\varphi\in\E$ defines a vector $\widetilde\varphi$ belonging to
    $\int_{[\nu_1,\infty)}^\oplus\de\lambda\,\H(\lambda)$. Thus, using partial integration for the
    terms involving derivatives with respect to $\lambda$, one finds that
    \begin{equation*}
      \left|\<(1\otimes Q^3)\psi,\varphi\>\right|=\left|\<\U\psi,\widetilde\varphi\>\right|
      \leq\textrm{Const.}\:\!\|\psi\|
    \end{equation*}
    for all $\psi\in\ltwo(\Sigma)\odot\mathscr S(\R)$, $\varphi\in\E$. Since
    $(1\otimes Q^3)\upharpoonright\ltwo(\Sigma)\odot\mathscr S(\R)$ is essentially selfadjoint, this
    implies that $\varphi\in\dom(1\otimes Q^3)$.
    
    (b) By Proposition \ref{S diff} the function $\lambda\mapsto S(\lambda)$ is three times
    continuously norm differentiable. Thus the argument in point (a) with $\varphi$ replaced by
    $S\varphi$ gives the result.
  \end{proof}

  \begin{Theorem}\label{short-range theorem}
    Let $H:=H_0+V$, where $V$ satisfies Assumption \ref{assumption V} with $\kappa>4$. Then, for each
    $\varphi\in\E$, $\tau_r(\varphi)$ exists for all $r>0$ and $\tau_r(\varphi)$ converges as
    $r\to\infty$ to a finite limit equal to $\<\varphi,\tau_\textsc{e-w}\varphi\>$.
  \end{Theorem}

  \begin{proof}
    We apply Theorem \ref{abstract time delay}. The hypotheses 1 and 2 of that theorem are satisfied
    due to Corollary \ref{wave operators}, and the hypotheses on $\varphi\in\E$ follow from Lemmas
    \ref{l1 integrals} and \ref{inclusions}. Since the function $\lambda\mapsto S(\lambda)$ is
    strongly continuously differentiable on $(\nu_1,\infty)\setminus\(\sigma_{\rm p}(H)\cup\T\)$,
    the proof is complete.
  \end{proof}
  
  \section*{Appendix}

  \begin{proof}[Proof of Lemma \ref{diff. of trace}]
    (a) Fix $\lambda\in(\nu_1,\infty)\setminus\T$ and let
    $\varphi\in\ltwo(\Sigma)\odot\mathscr S(\R)$. Choose $f\in C^\infty_0(\R)$ such that
    \begin{equation*}
      \big[1\otimes\gamma(\pm\sqrt{\lambda-\nu_\alpha})\big]\varphi=
      \big[1\otimes\gamma(\pm\sqrt{\lambda-\nu_\alpha})\big][1\otimes f(Q)]\varphi
    \end{equation*}
    for each $\alpha\in\N(\lambda)$. Then we get
    \begin{align*}
      \|T(\lambda)\varphi\|^2_{\H(\infty)}\leq\textrm{Const.}
      \sum_{\alpha\in\N(\lambda)}
      \Big\{\Big\|\big[&1\otimes\gamma(-\sqrt{\lambda-\nu_\alpha})f(Q)\big]
      \varphi\Big\|^2_{\ltwo(\Sigma)}\\
      &+\Big\|\big[1\otimes\gamma(\sqrt{\lambda-\nu_\alpha})f(Q)\big]
      \varphi\Big\|^2_{\ltwo(\Sigma)}\Big\}\:\!.
    \end{align*}
    Since $\gamma(\pm\sqrt{\lambda-\nu_\alpha})$ extends to an element of $\B\(\H^s(\R),\C\)$
    \cite[Thm.~2.4.2]{Kuroda78} and $f(Q)$ is bounded from $\H^s_t(\R)$ to $\H^s(\R)$,
    this implies that
    \begin{equation*}
      \|T(\lambda)\varphi\|^2_{\H(\infty)}=\textrm{Const.}\:\!
      \|\varphi\|^2_{\ltwo(\Sigma)\otimes\H^s_t(\R)}\:\!.
    \end{equation*}
    
    (b) Let $K$ be a compact set in $(\nu_1,\infty)\setminus\T$. Choose $\delta=\delta(K)>0$ such
    that $\lambda_1$ and $\lambda_2$ belong to the same interval $(\nu_\alpha,\nu_{\alpha+1})$
    whenever $\lambda_1,\lambda_2\in K$ and $|\lambda_1-\lambda_2|<\delta$. Let
    $\varphi\in\ltwo(\Sigma)\odot\mathscr S(\R)$. Due to the point (a), it is enough to show that
    there exist $\zeta>0$ such that
    \begin{equation}\label{sufficient estimate}
      \|\[T(\lambda_1)-T(\lambda_2)\]\varphi\|_{\H(\infty)}\leq\mathrm{Const.}\:\!
      |\lambda_1-\lambda_2|^\zeta\:\!\|\varphi\|_{\ltwo(\Sigma)\otimes\H^s_t(\R)}
    \end{equation}
    if $\lambda_1,\lambda_2\in K$ and $|\lambda_1-\lambda_2|<\delta$.

    Choose $f\in C^\infty_0(\R\setminus\{0\})$ such that
    \begin{equation*}
      (\lambda-\nu_\alpha)^{-1/4}\big[1\otimes\gamma(\pm\sqrt{\lambda-\nu_\alpha})\big]\varphi=
      \big[1\otimes\gamma(\pm\sqrt{\lambda-\nu_\alpha})\big]
      \big[1\otimes|Q|^{-1/2}f(Q)\big]\varphi
    \end{equation*}
    for each $\lambda\in K$, $\alpha\in\N(\sup K)$. Then we get
    \begin{align*}
      &\|[T(\lambda_1)-T(\lambda_2)]\varphi\|^2_{\H(\infty)}\\
      &\leq\textrm{Const.}\sum_{\alpha\in\N(\lambda_1)}
      \Big\{\left\|1\otimes\big[\gamma(-\sqrt{\lambda_1-\nu_\alpha})
	-\gamma(-\sqrt{\lambda_2-\nu_\alpha})\big]
      \big[1\otimes|Q|^{-1/2}f(Q)\big]\varphi\right\|^2_{\ltwo(\Sigma)}\\
      &\qquad\qquad\qquad\qquad
      +\left\|1\otimes\big[\gamma(\sqrt{\lambda_1-\nu_\alpha})
	-\gamma(\sqrt{\lambda_2-\nu_\alpha})\big]
      \big[1\otimes|Q|^{-1/2}f(Q)\big]\varphi\right\|^2_{\ltwo(\Sigma)}\Big\}\:\!.
    \end{align*}
    Since the function $\R\ni\xi\mapsto\gamma(\xi)\in\B\(\H^s(\R),\C\)$ is H\"older continuous
    \cite[Thm.~2.4.2]{Kuroda78} and $|Q|^{-1/2}f(Q)$ is bounded from $\H^s_t(\R)$ to
    $\H^s(\R)$, this implies (\ref{sufficient estimate}).
    
    (c) The proof is similar to that of \cite[Lemma~3.3]{Jensen81}.
  \end{proof}

  \begin{proof}[Proof of the embedding $\ltwo(\Sigma)\otimes\H^{-1}_t(\R)\subset\K$ for any
      $t>1/2$]
    Since $\ltwo(\Sigma)\otimes\H^{-1}(\R)\subset\G^{-1/2}$ and
    $\dom(A;\G^{-1/2})\subset\G^{-1/2}\cap\dom(A;\G^{-1})$, we have
    $\(\dom[A;\ltwo(\Sigma)\otimes\H^{-1}(\R)],\ltwo(\Sigma)\otimes\H^{-1}(\R)\)_{1/2,1}\subset\K$
    due to \cite[Cor.~2.6.3]{ABG}. Then we obtain that
    $\(\dom[A;\ltwo(\Sigma)\otimes\H^{-1}(\R)],\ltwo(\Sigma)\otimes\H^{-1}(\R)\)_{\mu,2}\subset\K$
    for any $\mu<1/2$, by using \cite[Thm.~3.4.3.(a)]{ABG}. Since
    $\ltwo(\Sigma)\otimes\H^{-1}_1(\R)\subset\dom[A;\ltwo(\Sigma)\otimes\H^{-1}(\R)]$, this leads
    to the embedding
    $\(\ltwo(\Sigma)\otimes\H^{-1}_1(\R),\ltwo(\Sigma)\otimes\H^{-1}(\R)\)_{\mu,2}\subset\K$
    \cite[Cor.~2.6.3]{ABG}. Now, by using \cite[Thm.~12.6.1]{Aubin00} and
    \cite[Thm.~VII(I.1)]{Lions/Peetre}, we get the isometry
    $\ltwo(\Sigma)\otimes\H^{-1}_{1-\mu}(\R)\simeq
    \(\ltwo(\Sigma)\otimes\H^{-1}_1(\R),\ltwo(\Sigma)\otimes\H^{-1}(\R)\)_{\mu,2}$. Therefore
    $\ltwo(\Sigma)\otimes\H^{-1}_t(\R)\subset\K$ for any $t>1/2$.
  \end{proof}

  \section*{Acknowledgements}

  We thank S. Richard for having pointed out to us similarities between $N$-body scattering and
  scattering in waveguides. We are also grateful to W. O. Amrein and A. Jensen for their helpful
  remarks. This work was partially supported by the Swiss National Science Foundation.

  \providecommand{\bysame}{\leavevmode\hbox to3em{\hrulefill}\thinspace}
  \providecommand{\MR}{\relax\ifhmode\unskip\space\fi MR }
  \providecommand{\MRhref}[2]{%
    \href{http://www.ams.org/mathscinet-getitem?mr=#1}{#2}
  }
  \providecommand{\href}[2]{#2}
  
  
\end{document}